# Automated Seizure Detection: Unrecognized Challenges, Unexpected Insights

Ivan Osorio[1], Alexey Lyubushin[2], Didier Sornette[3]

[1]Corresponding Author
Comprehensive Epilepsy Center
University of Kansas Medical Center
3901 Rainbow Blvd.
Kansas City, KS 66209
iosorio@kumc.edu
913 5884529
913 5884585 (fax)

[2]Institute of Physics of the Earth, Russian Academy of Sciences
B.Gruzinskaya, 10
Moscow, Russia, 123995

[3]ETH Zurich
 Department of Management, Technology and Economics
Kreuzplatz 5, CH-8032
Zurich, Switzerland

**Abstract**

One of epileptology's fundamental aims is the formulation of a universal, internally consistent seizure definition. To assess this aim's feasibility, three signal analysis methods were applied to a seizure time series and performance comparisons were undertaken among them and with respect to a validated algorithm. One of the methods uses a Fisher's matrix weighted measure of the rate of parameters change of a $2^{nd}$ order auto-regressive model, another is based on the Wavelet Transform Maximum Modulus for quantification of changes in the logarithm of the standard deviation of ECoG power and yet another employs the ratio of short-to-long term averages computed from cortical signals.

The central finding, *fluctuating concordance* among all methods' output as a function of seizure duration, uncovers unexpected hurdles in the path to a universal definition, while furnishing relevant knowledge in the dynamical (spectral non-stationarity and varying ictal signal complexity) and clinical (probable attainability of consensus) domains.

**Highlights**

Consensus among epileptologists as to what grapho-elements are classifiable as ictal, is difficult to achieve.

Adoption of a universal seizure definition would be of heuristic value.

Four signal processing methods were applied to a seizure time seizures to identify ictal markers.

Concordance among the various methods for key metrics such as sensitivity, specificity and speed of detection varied as a function of seizure duration.

Discordance among methods hints at fluctuating complexity/entropy of ictal signals and potential unattainability of a universal, internally consistent seizure definition.

**Introduction**

Real-time ("on the run") automated seizure detection provides the only means through which contingent warning to minimize risk of injury to patients or delivery of a therapy for control of seizures may be implemented. Performance assessment of these algorithms relies entirely on expert visual analysis, which provides the benchmarks (seizure onset and end times) from which key metrics (detection latency in reference to electrographic and clinical onset time ("speed of detection"), sensitivity, specificity and positive predictive value) are derived. To date, performance comparisons among myriad existing algorithms have not been performed due to lack of a common and adequate database, a limitation that is being addressed (see Schulze-Bonhage article in this issue). However, if and when undertaken, said "comparisons" would be largely unwarranted and have meager, if any, clinical value/translatability, given that no universally accepted definition of what is a "seizure" has been crafted [1-3]. The process of evaluation of seizure detection algorithms is plagued with cognitive biases and other confounding intricacies [4,5] that impede achievement of consensus and in certain cases even of majority agreement [6].

 The main purpose of this investigation is to assess the performance of signal analysis methods that rely principally on power variance to those that rely mainly on power spectral shape for detection of seizures. These inter-related features (power variance and spectral shape) are the subject of study since they are exploited by the majority of existing seizure detection algorithms. The results support the usefulness of the various methods for seizure detection, revealing differences that may be substantive, but also enlightening. The implications of said differences on the prospect for transcending expert visual analysis subjectivity through the crafting of an objective, universally acceptable definition and the importance of alternative approaches are analyzed and discussed. An article (Osorio et al., this issue) presents a strategy to address ostensible human and cybernetic "inconsistencies" about which changes in cortical electrical activity constitute a seizure.

**Methods**

**Seizure Detection Methods**

   The following signal analysis methods were applied to the electrocorticogram (ECoG) to derive metrics for the discrimination of seizure from non-seizure signals:

   (i) An Auto-Regression (AR) model of the 2$^{nd}$ order, yielding autoregression coefficients and the logarithm of residual variance,

   (ii) Estimates of the logarithm of the standard deviation (SD) of differentiated ECoG using long chains of wavelet transform modulus maxima (WTMM chains) based on the first derivative of a Gaussian function $\sim \exp(-t^2)$ as a continuous wavelet kernel, and

   (iii) The ratio of the "short time average" (STA) to the "long time average" (LTA), widely used in seismology for precise real-time earthquake detection [7]. The spectral and dynamical similarities between seizures and earthquakes [8,9] provide the motivation for application of this method to epileptology.

   (iv) A validated seizure detection algorithm [6,10] is used as a reference to better interpret the results of the novel proposed ones and to cast light on the intricacies and challenges of discriminating seizure from non-seizure signals even when using objective, quantitative means.

**Human Seizure Time Series/ECoG**

Data obtained from one subject undergoing evaluation for epilepsy surgery with intra-cranial electrodes was selected for analyses as it had the largest number of clinical and subclinical seizures in the University of Kansas Medical Center Epilepsy Database. ECoG was collected in accordance with the Center's surgical evaluation protocol and with the Human Subjects Committee requirements, which include signing of a consent form by the subject.

The ECoG was recorded using electrodes implanted into the amygdala, pes hippocampus and body of hippocampus bilaterally through the temporal neocortex and had a duration of 6.9 days (142'923'853 samples; 239.75 Hz sampling rate).

**Differentiation of the ECoG signals used in the analyses**

For efficient analyses, ECoG signal differentiation was performed, so as to minimize the non-stationarity present in them. If $Z(t)$ is raw ECoG, then its difference is $X(t) = Z(t) - Z(t-1)$, where (t) corresponds to a sample time increment. This linear operation is exactly invertible and, unlike band-pass filtering or detrending, does not suppress low frequency fluctuations, but decreases their overall influence. Figure 1 illustrates the effect of this operation on raw ECoG and figure 2 shows a time-frequency map of the evolution of the power spectra of differentiated ECoG segments. The power spectra are estimated within 5 sec moving windows of length.

**Signal Analysis Methods**

**Autoregression Model or " $r^2$ -Method"**

The autoregression model ( $r^2$ -method) [11-13] is estimated on differentiated ECoG in a doubly moving time window, that is, a sliding window consisting of a center and two parts of equal length, one to the right (the foreground or "future" signals) and the other to the left of its center (the background or "past" signals). This approach increases the precision with which onset and termination times are determined. The AR-model is estimated within the "left" and "right" window halves and the distance between the vectors of AR-parameters is calculated using the Fisher matrix. This method may be regarded as operating in "real-time" as its detection delay equals the length of the foreground half-window (1 sec.).

The method's parameters (all adaptable) used in this application are: a). AR-order (p) = 2; b) Length (L) of the half moving time window = 1sec, and c) Detection Threshold (R) = 3, for specifying seizure onsets and termination times. Given the half-window length $L = 240$ and the data sampling rate (240 Hz), the delay inherent in identifying a seizure onset or termination is equal to $5L/4$ or 1.25 sec (see rule 6 below).

**Seizure Detection with the " $r^2$ -Method"**

Consider the autoregressive model of the $p$ -th order [12,13] for the signal increments:

$$X(t) + \sum_{k=1}^{p} a_k \cdot X(t-k) = d + \varepsilon(t), \quad M[\varepsilon(t)] = 0, \quad M[\varepsilon^2(t)] = \sigma^2 \qquad (1)$$

where $M[.]$ is the symbol for a mathematical expectation. The model (1) can be re-written in a more compact form:

$$X(t) = c^T Y(t) + \varepsilon(t), \quad Y(t) = (-X(t-1), ..., -X(t-p), 1)^T, \quad c = (a_1, ..., a_p, d)^T \qquad (2)$$

where $c^T Y(t)$ is a scalar product of column-vectors with $c^T$ being the transposed vector of $c$. Thus, in the AR-model, each sample is presented as a weighted sum of $p$ previous values with weights given by the AR-coefficients, plus some shift $d$ and the residual $\varepsilon(t)$, which is regarded as noise with zero mean value and variance $\sigma^2$. The full vector of parameters of the AR-model is $\theta = (c^T, \sigma)^T$.

A vector $\theta$ is estimated for each of the two moving time half-windows of equal length $L$ to the left (background) and right (foreground) of the window's center $\tau$. Let $\theta^{(1)}$ be the left half-window parameter vector and $\theta^{(2)}$ the right half-window parameter vector, and $\Delta\theta = \theta^{(2)} - \theta^{(1)}$ their difference. Their difference is weighed using Fisher's matrix for the model (1) defined by:

$$B = -\frac{\partial^2 \ln(\Phi)}{\partial\theta\,\partial\theta}, \quad \ln(\Phi) = -(L-p)\ln(\sigma) - \frac{1}{2\sigma^2}\sum_t (X(t) - c^T Y(t))^2 \qquad (3)$$

Expression (3) defines $B$ as the matrix constructed from the second-order derivatives of the logarithm of the likelihood function $\Phi$ under the assumption that $\varepsilon(t)$ is Gaussian white noise. Let $B^{(1)}$ and $B^{(2)}$ be the matrices (3) computed in the left and right halves of the moving time window and let us introduce a measure of non-stationarity:

$$r^2(\tau) = (\Delta\theta^T B^{(1)} \Delta\theta + \Delta\theta^T B^{(2)} \Delta\theta)\,/(2(L-p)) \qquad (4)$$

This measure [14] provides a natural dimensionless estimate of the non-stationary behavior of the signal $X(t)$. To make the calculation explicit, this equation (4) is estimated by using the following expression:

$$\Delta\theta^T B \Delta\theta = \frac{2(\Delta\sigma)^2}{\sigma^2} + \frac{\Delta c^T (\sum_t Y(t) Y^T(t)) \Delta c}{\sigma^2 (L-p)} + \frac{4\Delta c^T \Delta\sigma \sum_t \varepsilon(t) Y(t)}{\sigma^3 (L-p)} \qquad (5)$$

The non-stationarity measures (4)-(5) will be used to identify the onset and termination of seizures based on the condition that a local maximum of $r^2$ exceeds a given threshold $R$. Specifically, if

$$r^2(\tau) = \max_{\xi} \{r^2(\xi),\ \tau - L/4 \le \xi \le \tau + L/4\}, \quad r^2(\tau) \ge R, \qquad (6)$$

then, the time $\tau$ is

1.  the onset of a seizure if $\sigma_2 > \sigma_1$ (the variance of the residuals of the AR process is larger in the right half of the window (foreground) than in the left half (background);
2.  the end of a seizure if $\sigma_1 > \sigma_2$ (the variance of the residuals of the AR process is smaller in the right half of the window than in the left half).

Condition (6) reflects the large non-stationarity present in the signal associated with the onset or end of seizures as determined by the jumps from low to high variance (onset) or vice-versa (end) at time $\tau$.

The values of the residual variances $\sigma_1$ and $\sigma_2$ are the parameters of the "$r^2$-Method" as well as components of the vector $c$, which is why they are consolidated into a general vector of parameters $\theta = (c^T, \sigma)^T$. The method is based on comparing vectors $\theta_1$ (left half-window) and $\theta_2$ (right-half window) using Fisher's matrix as a "natural" statistical metric. It is worth pointing out that the AR(2) method is not sensitive to changes of variance in power, but to changes in the *shape* of the spectral density; this is because a short time window estimate $\hat{S}_{XX}(\omega)$ of the spectral density is directly connected with the vector of parameters $\theta$ by the equation $\hat{S}_{XX}(\omega) = \sigma^2 / (2\pi \cdot |1 + a_1 e^{-i\omega} + a_2 e^{-2i\omega}|^2)$ [11], where $\omega$ is an angular frequency and $i$ is the imaginary unit. This connection makes this method sensitive to changes in the auto-covariance function $R_{XX}(k) = M\{X(t)X(t-k)\}$ as follows from the Wiener–Khinchin theorem:

$$S_{XX}(\omega) = \sum_{k=-\infty}^{+\infty} R_{XX}(k) \cdot e^{-ik\omega}, \text{ where } R_{XX}(k) = \int_{-\pi}^{\pi} S_{XX}(\omega) \cdot e^{ik\omega} d\omega.$$

### Continuous Wavelet Transform Maximum Modulus (WTMM)

This method exploits the large, sudden changes in the variance of signal features, such as amplitude and frequency that characterize most epileptic seizures. These features are calculated within a hierarchy of sequences of short contiguous windows of equal length and are extracted in the form of chains of continuous Wavelet Transform Maximum Modulus (WTMM), using a mother wavelet defined as the 1st order derivative of the Gaussian function $\sim \exp(-t^2)$. The construction of long chains to form a WTMM-skeleton is a commonly used method for detecting changes in the mean value of noisy signals such as edge detection in vision computer programs. WTMM relies on three parameters: a) The length of the window used to calculate the sequence of variance values; b) A parameter $a_*$ that defines which WTMM-chains are long and, c) The threshold $_{\Delta U}$ for the stepwise approximation of the logarithmic variance curve for detecting the onsets and ends of seizures.

The Wavelet Transform may be conceptualized as a "mathematical microscope" that is well suited to reveal the hierarchy that governs the spatial distribution of singularities of multifractal measures ([15-17]. The role of the scaling parameter $a_*$ is similar to that of the magnification setting in a microscope: the larger the value of $a_*$, the larger the scale of the signal's structure under investigation. By using wavelets instead of boxes, the smoothing effect of polynomials that might either mask singularities or perturb the estimation of their strength (Hölder exponent) is avoided. This approach remedies one of the main failures of the classical multifractal methods such as the box-counting algorithms in the case of measures and of the structure function method in the case of functions [18-21]. Another advantage of this method is that the skeleton defined by the WTMM [22,23] provides an adaptive space-scale partitioning from which to extract the singularity spectrum via the Legendre transform of the scaling exponents (real, positive as well as negative) of some partition functions defined from the WT skeleton. The reader is referred to Bacry et al., [19] and Jaffard [24,25] for rigorous mathematical description and applications and to Hentschel [26] for the theoretical treatment of random multifractal functions.

As originally pointed out for the specific purpose of analyzing the regularity of a function [22,23], the redundancy of the WT may be eliminated by focusing on the WT skeleton defined by its modulus maxima only. These maxima are defined, at each scale $a$, as the local maxima of the set of wavelet transforms over all possible position $x$ at the fixed scale $a$ of the function $f$ denoted $|T_\psi[f](x,a)|$. These WTMM points are disposed on connected curves in the space-scale (or time-scale) half-plane, called maxima lines. Let us define $L(a_0)$ as the set of all the maxima lines that exist at the scale $a_0$ and which contain maxima at any scale $a < a_0$. An important feature of these maxima lines, when analyzing singular functions, is that there is at least one maxima line pointing towards each singularity [21-23].

There are almost as many analyzing wavelets as applications of the continuous WT [15-21] [22,23]. A commonly used class of analyzing wavelets is defined by the successive derivatives of the Gaussian function:

$$g^{(N)}(x) = \frac{d^N}{dx^N} \exp(-x^2/2) \tag{7}$$

Note that the WT of a signal $f(x)$ with $g^{(N)}$ (Eq. 7) takes the following simple expression:

$$T_{g^{(N)}}[f](x,a) = \frac{1}{a} \int_{-\infty}^{+\infty} f(y) g^{(N)}((y-x)/a) dy = a^N \frac{d^N}{dx^N} T_{g^{(0)}}[f](x,a) \tag{8}$$

Equation (8) shows that the WT computed with $g^{(N)}$ at scale $a$ is nothing but the N-th derivative of the signal $f(x)$ smoothed by a dilated version $g^{(0)}(x/a)$ of the Gaussian function. This property is at the heart of various applications of the WT "microscope" as a very efficient multi-scale singularity tracking technique [27].

**Algorithm to construct the WTMM-skeleton**

Let $Y(t)$ be an arbitrary signal and consider the corresponding smoothed signal obtained by using a scale-dependent kernel:

$$c_0(t,a) = \int_{-\infty}^{+\infty} Y(t+av) \cdot \psi_0(v) \, dv \Big/ \int_{-\infty}^{+\infty} \psi_0(v) \, dv \tag{9}$$

where $a > 0$ is a time scale and $\psi_0(t)$ is some function decaying sufficiently fast on both sides of its single maximum; further on, $\psi_0(t) = \exp(-t^2)$ shall be used. The wavelet function is defined as:

$$\psi_n(t) = (-1)^n \cdot \frac{d^n \psi_0(t)}{dt^n} \equiv (-1)^n \cdot \psi_0^{(n)}(t) \tag{10}$$

Using integration by part and exploiting the fast decay of the function $\psi_0(t)$ at $t \Rightarrow \pm\infty$, the following formula for the Taylor's coefficients (the $n$-th derivative of the smoothed signal, divided by $n!$) of the smoothed signal is obtained:

$$c_n(t,a) \equiv \frac{1}{n!}\frac{d^n c_0(t,a)}{dt^n} = \int_{-\infty}^{+\infty} Y(t+av)\psi_n(v)\,dv \bigg/ a^n \int_{-\infty}^{+\infty} v^n \psi_n(v)\,dv \qquad (11)$$

Equation (9) is a particular case of formula (11) for $n=0$.

The WTMM-point $(t,a)$ for $n \geq 1$ is defined as the point for which $|c_n(t,a)|$ has a local maximum with respect to time $t$ for a given time scale $a$ [26]. For $n=0$, the WTMM-points are defined as points of local extremes (maxima or minima) of the smoothed signal $c_0(t,a)$ that may be joined to form chains; the set of all chains creates a WTMM-skeleton of the signal. If $\psi_0(t)$ is a Gaussian function, then a given WTMM-skeleton chain does not end when the scale $a$ is decreased [29]. The WTMM-points for the 1$^{st}$ order derivative $c_1(t,a)$ indicate time points of the maximum trend (positive or negative) of the smoothed signal $c_0(t,a)$ for the given scale. This allows temporal localization of "points" of large and abrupt changes of the mean value of a noisy signal such as ECoG; these "points" mark the times when rather long WTMM-chains begin to grow.

A stepwise approximation $S_Y(t\,|\,a_*)$ for the signal $Y(t)$ is defined as a function that is equal to the sequence of constant values $g_k$ in the successive intervals $t \in [\tau_k(a_*), \tau_{k+1}(a_*)]$. Here $\tau_k(a_*)$ are the beginnings of the WTMM-chains for $c_1(t,a)$ that exceed the threshold time scale $a_*$ and

$$g_k = \sum_{t=\tau_k(a_*)}^{\tau_{k+1}(a_*)} Y(t)/(\tau_{k+1}(a_*) - \tau_k(a_*) + 1) \qquad (12)$$

is equal to the mean value of $Y(t)$ within the time interval $[\tau_k(a_*), \tau_{k+1}(a_*)]$.

For example, consider a signal that is a sum $Y(t) = S_0(t) + \varepsilon(t)$, $t = 1,...,2000$, where $\varepsilon(t)$ is a Gaussian white noise with unit standard deviation and
$S_0(t) = 0$ for $t \in [1,250],[751,1250],[1751,2000]$;
$S_0(t) = 2$ for $t \in [251,750],[1251,1750]$.

The resulting stepwise approximation for this signal using WTMM-chains is shown in Figure 3. Thus, times of abrupt changes of the mean value of the signal of interest can be estimated as instants of rather large steps of the signal $S_Y(t\,|\,a_*)$ with an appropriate choice of the time scale threshold $a_*$.

## 2.1 Seizure Detection with the WTMM-method

The WTMM-method of constructing a stepwise approximation $S_Y(t\,|\,a_*)$ for an arbitrary signal $Y(t)$ may be applied to the task of seizure detection. This is suggested from figures 1 and 2, which show seizures as intervals of large, abrupt signal variance. Let us calculate sample estimates of the variance of the normalized ECoG $X(t)$ within "small" adjacent time intervals of length $L$ and take the logarithmic values of the standard deviations:

$$U(\xi_j) = \tfrac{1}{2} \cdot \lg(V_X(\xi_j)), \quad V_X(\xi_j) = \tfrac{1}{L} \sum_{t=\xi_j - L + 1}^{\xi_j} X^2(t), \quad \xi_j = j \cdot L, \quad j = 1, 2, \ldots \qquad (13)$$

Thus, $U(\xi_j)$ is a time series with a sampling time interval that is $L$ times larger than the sampling time interval of the original ECoG. Figure 4 shows $U$ plotted as a function of the time position of the right end of these time intervals of length $L$ as well as its stepwise approximation $S_U(\xi_j \mid a_*)$, where $a_*$ is a scale threshold for the decimal logarithms of the variance $U$ of the ECoG.

Seizure onsets and terminations are defined by the following rule:

$$\begin{aligned} \xi_j \text{ is onset if} \quad & S_U(\xi_{j+1} \mid a_*) - S_U(\xi_j \mid a_*) \geq \Delta U \\ \xi_j \text{ is end if} \quad & S_U(\xi_{j+1} \mid a_*) - S_U(\xi_j \mid a_*) \leq \Delta U \end{aligned} \qquad (14)$$

The time interval length $L$, the scale threshold $a_*$ for the decimal logarithms of the variance and the threshold value $\Delta U$ for the logarithm of the variance of the normalized ECoG are this method's three parameters. In the application shown in figure 4, the chosen parameter values are the following: $(L = 240, a_* = 3, \Delta U = 0.25)$. The value of the threshold $\Delta U$ determines the "strength" required to detect ECoG activity that qualifies as seizures. Given a window $L = 240$ for calculating the sequence of $U(\xi_j)$ and a sampling rate of 240 Hz, the time resolution of the WTMM-seizures detection implementation shown in figure 4 is 2 s.

**Seizure Detection with the STA/LTA method.**

The "short time average" (STA) divided by the "long time average" (LTA) is widely used in seismology as an earthquake detector and has several realizations [7], one of which will be used here. The dynamical analogies between earthquakes and seizures provide a rationale for applying this method to their detection [8,9].

Let $X(t)$ be the output of the Daub04 3$^{\text{rd}}$ level band-pass filter applied to the ECoG, $N_{STA}$ the length (in number of samples) of the "short time average" and let $N_{LTA}$ be the length (in number of samples) of the "long time average". The STA/LTA ratio is defined by the formula:

$$STALTA(\tau) = \frac{\sum_{t=\tau - N_{STA} + 1}^{\tau} X^2(t) \Big/ N_{STA}}{\sum_{t=\tau - N_{LTA} + 1}^{\tau} X^2(t) \Big/ N_{LTA}} \qquad (15)$$

where $\tau$ is the common right-hand end of both short and long averaging time windows.

Seizure onsets $\tau_{onset}$ correspond to the times for which the following condition is fulfilled:

$$STALTA(\tau + N_{STA}/2) \geq T_{onset} \qquad (16)$$

Seizure terminations $\tau_{end}$ are determined using the rule $\tau_{end} = \tau^* - N_{STA}/2$ where $\tau^*$ is the time $\tau$ after the most recent seizure onset for which the following condition is fulfilled:

$$\max\{\, STALTA(s),\, \tau - N_{STA} \leq s \leq \tau \,\} \leq T_{end} \qquad (17)$$

The STA/LTA-detector has the following parameters: 1. The length of the short time average $N_{STA}$; 2. the length of the long time average $N_{LTA}$; 3. the threshold $T_{onset}$ for seizure onset as defined above; 4. the threshold $T_{end}$ for seizure termination as defined above; their values for this application were chosen to be: $N_{STA} = 360$, $N_{LTA} = 16 \cdot N_{STA}$ and $T_{onset}$ and $T_{end}$ values were chosen under the condition $T_{onset} > T_{end}$. Using different threshold values, it is possible to increase or decrease the temporal resolution with which seizures are detected.

### 3. Seizure Detection with a Validated Algorithm (*Val*)

This algorithm [6,10] recognizes seizures by their rapid increase in signal (ECoG) power in a particular weighted frequency band (8-42 Hz). In its generic embodiment, it operates with the following (adaptable) parameters: 1. A threshold value, $T$ (equal to 22); 2. A duration constraint $D$ (taken equal to 0.84 s). Specifically, this algorithm continuously calculates a dimensionless ratio in moving windows by dividing the signal power estimated over a foreground window of 2 s. duration by that of a background window of 30 min. so as to estimate the signal's seizure content. Based on detailed and rigorous analyses of long time series recorded from humans with epilepsy, it has been determined that a dimensionless ratio reaching a value $T = 22$ for at least 0.84 s. has a greater than 90% probability of corresponding to a seizure onset. These threshold and duration constraints were chosen to optimize specificity of detection through the elimination of bursts of epileptiform discharges lasting less than 0.84s in conformance with the intended application (warning and therapy delivery), which for certain applications requires high specificity.

Let $X(t)$ be the output of the ECoG after "passing" through the Daubechies 04, 3$^{rd}$ level wavelet acting as a band-pass filter, $Y^2(t) = X(t)$. Let $Q$ be the length of a moving time window and $\tau$ be the number of samples. The foreground ($FG$) is defined as the median value:

$$FG(\tau) = median\{Y(s),\, \tau - Q + 1 \leq s \leq \tau\} \qquad (18)$$

whereas the background ($BG$) is defined using another median value:

$$BG(\tau) = median\{Y(s_k),\, s_k = \tau - (k-1)\Delta s, k = 1, ..., Q\},\, \Delta s > 1 \qquad (19)$$

The value (19) is computed for discrete values of $\tau$, taken with step $\Delta s$: $\tau = \tau_j$, $\tau_{j+1} = \tau_j + \Delta s$. With these definitions, the background is defined according to the formula:

$$BG^{(\lambda)}(\tau) = \begin{cases} (1-\lambda) \cdot BG(\tau) + \lambda \cdot BG^{(\lambda)}(\tau-1), & \tau = \tau_j \\ BG^{(\lambda)}(\tau-1), & \tau_j < \tau < \tau_{j+1} \end{cases} \qquad (20)$$

Finally, a value of the ratio:

$$R^{(\lambda)}(\tau) = FG(\tau) / BG^{(\lambda)}(\tau) \qquad (21)$$

is set for detecting seizures  A seizure detection is issued when $R^{(\lambda)}(\tau) \geq T$ and terminated when

$$\tau = \tau_{end}: \quad R^{(\lambda)}(\tau) \leq T, \quad \tau \geq \tau_{onset} + D \qquad (22)$$

The main parameters are $(Q, \Delta s, \lambda, T)$. Their values in the *Val* method are the following: $Q = 480$, $\Delta s = 120$, $\lambda = 0.999807$, $T = 22$. The use of order statistics (median given by expression (19) in this embodiment) and a slowly decaying ($\lambda$) exponential memory with average decay time steps for the background window provide stability and robustness to the ratio (21) with respect to the influence of single epileptiform discharges or short bursts. All parameters are adaptable so as to optimize performance for the application at hand.

### Results

The total number of detections, their duration and the percent time spent in seizure over the time series total duration (6.9 days) are presented in Table 1. The STA/LTA yielded the largest number of detections, but only the third largest time spent in seizure, given the shortness of median duration of detection compared to those computed by the WTMM and $r^2$ methods. The mean and median durations of detections issued by the $r^2$ method were the longest, but the WTMM algorithm surpassed all others in duration of time spent in seizure.  Figures 5a, 5b and 5c illustrate the differences between the four methods.

In order to better understand these differences, an indicator function (IF) is applied to the results. IF equals 1 for the duration of a seizure and 0 before its onset and after its termination (0 corresponds to non-seizure intervals as identified by each method). The calculation of the IF generates four-stepwise time functions, one for each detection method: $\chi_{Val}(t)$, $\chi_{r^2}(t)$, $\chi_{STA/LTA}(t)$ and $\chi_{WTMM}(t)$. Using this IF, two additional functions are computed over a 0.1 s running window: a) The average indicator function (*AIF*):

$$AIF(t) = (\chi_{Val}(t) + \chi_{r^2}(t) + \chi_{STA/LTA}(t) + \chi_{WTMM}(t)) / 4 \qquad (23)$$

and b) The product of indicator functions (*PIF*):

$$PIF(t) = \chi_{Val}(t) \cdot \chi_{r^2}(t) \cdot \chi_{STA/LTA}(t) \cdot \chi_{WTMM}(t) \qquad (24)$$

The *AIF*'s values may vary between $[0-1]$ (and can take on any intermediate value 0.25, 0.5, 0.75 in this application) whereas the *PIF* values are either 0 or 1; a $PIF = 1$ corresponds to an $AIF = 1$ and a $PIF = 0$, to an $AIF < 1$. Time intervals for which $AIF = PIF = 1$ correspond to seizures detected by all methods. Inspection of Figure 5a–c reveals that AIF values are smaller than 1 (e.g., only one or two out of the four methods recognize the activity as ictal in nature) at the onset and termination of certain type of ECoG activity but frequently reach 1, sometime into the ictus, as all methods "reach consensus". Table 2 provides further evidence that, at some point in time, the majority of seizures detected by the validated algorithm are also detected by the other three methods, with WTTM detecting the largest number (97%) and STA/LTA the second largest (91.5%) number of seizures. More specifically and by way of example, the value 0.971 in Table 2 means that the WTMM method detections encompass 97.1% of seizure time intervals detected with the validated method, with the exception of 1.6 s. that correspond to the delay/lag between them in detecting seizure onsets (see below for details).

Time intervals for which the pairwise product $\chi_{Val}(t) \cdot \chi_{r^2}(t) = 1$ correspond to seizures detected by both the validated algorithm and $r^2$. Dividing the number of time intervals when $\chi_{Val}(t) \cdot \chi_{r^2}(t) = 1$ by the number of intervals when $\chi_{Val}(t) = 1$, yields the specificity of the $r^2$ method with respect to the validated algorithm. Since the validated algorithm has an inherent delay of 1 s (the median filter's foreground window is 2 s) and an additional duration constraint of 0.84 s. is imposed before a detection is issued, its onset and end times are "delayed" compared to those yielded by the other methods. To account for this delay and make comparisons more meaningful, the specificity of the $r^2$ with respect to the validated algorithm is re-calculated as a function of a time shift $\tau$ :

$$Spe_{r^2\_Val}(\tau) = \sum_t (\chi_{r^2}(t + \tau) \cdot \chi_{Val}(t)) \Big/ \sum_t \chi_{Val}(t) \qquad (25)$$

The specificity functions for the two other methods $Spe_{WTMM\_Val}(\tau)$ and $Spe_{STA/LTA\_Val}(\tau)$ are identically computed and their maximum value (dependent on $\tau$) may be regarded as the mean value of the time delay of one method's function with respect to another for seizure onset and end times. From the results shown in Figure 6, it can be seen that the time differences are negative for all methods with respect to the validated one; that is, the validated algorithm's detection times lag behind those given by the other methods. Particularly, the mean delay of the validated algorithm is 1.1 s with respect to $r^2$, 0.6 s with respect to STA/LTA and 1.6 s with respect to WTMM while the mean delay of $\chi_{Val}(t)$ with respect to $\chi_{PIF}(t)$ is 0 by construction. As expected, the re-calculated specificity values shifted by $\tau$ shown in Table 2 are higher compared to those without shifting.

Except for $Spe_{PIF\_Val}(\tau)$, the shape of the other specificity functions is asymmetric (Figure 6). Negative values of the specificity functions found for small positive mutual shifts $\tau$ are the consequence of the fact that, on average, these time shifts correspond to periods that the *Val* method does not classify as seizures, activity that the other methods do. This alternating effect for mutually shifted seizures time intervals is the strongest for the values of $\tau$ corresponding to the minimum of the cross-covariance functions. There are also instances when other methods do not classify some time intervals as seizures while the validated algorithm does.

The value $\max_{\tau} Spe_{PIF\_Val}(\tau) = Spe_{PIF\_Val}(0) = 0.554$ indicated in the lower right panel of figure 6 means that only 55.4% of seizures recognized as such by the other methods are also detected by the validated method, indicating that in its generic form and by design, it is less sensitive and more specific for seizure detection than the others.

**Discussion**

The three methods presented herein survey different but inter-dependent ECoG signal properties, thus expanding the breadth and perhaps also the depth of insight into the spectral "structure" of epileptic seizures in a clinically relevant manner. The Auto-Regressive model ($r^2$), sensitive mainly to changes in spectral shape, was chosen as the simplest and most general method, with which to provide a statistical description of oscillations (ECoG) that may be regarded as generated by the stochastic analogue of a linear oscillator. The *WTMM* method is well suited for estimations of changes in power variance within adjacent "short" time windows whereas the *STA/LTA* uses the ratio of variances to detect, at low computational expense, ECoG signal changes corresponding to seizures. The validated algorithm [6, 10] whose architecture is similar to that of the *STA/LTA* and is also sensitive to power variances within certain frequencies (8-45 Hz) was used as a "benchmark" since its performance has been subject to rigorous peer-review. The $r^2$ and *STA/LTA* algorithms are implementable into implantable devices as they operate in real-time, while the *WTMM* is best suited for off-line analysis applications given its relatively high algorithmic complexity.

Whereas various performance metrics for each algorithm pervade the Results section inevitably leading to comparisons among them, these would be misleading and misplaced given that each method not only operates with different parameters, but also probes *different* ECoG features. The discrepancies in number and duration of detections issued by each algorithm, which may be inherently or operationally "irreconcilable", parallel those that characterize and possibly define visual expert analysis ([6]; see Osorio et al., this issue). The fundamental implication of this observation is that a unified or universally acceptable "definition" of what activity constitutes a seizure may not be attainable (nor desirable) even through the application of objective, advanced signal analyses methods, particularly for seizure onset and termination segments. Algorithmic and visual expert analysis consensus as to what grapho-elements define a 'seizure' seems to be highly dependent on when during the course of a 'seizure' a decision is made. In this context, it is noteworthy that *AIF* and *PIF* frequently reached a value of 1, indicative of concordance among all detection methods sometime after seizure onset and before its termination (as determined by any of the methods), provided said seizures reached a certain duration (20-30s.) as it will be discussed in more detail in this issue's accompanying article. In short, seizure onsets and terminations may be under certain conditions universally *undefinable* by algorithmic or expert visual consensus. A systematic investigation of the differences in signal spectral properties between the "preface"/"epilogue" and the "main body" of seizures was not performed. It is speculated that the presence of "start-up transients" (in a dynamical sense) and of temporo-spatial dispersion of the ictal signal (which impacts S/N) may be most prominent at the onset and termination of seizures. These and local and global state-dependencies of certain signal features, account in part for the *temporal fluctuations* in algorithmic detection performance that characterize these results.

Defining seizure energy, as the product of the standard deviation of the power of ECoG by its duration (in seconds), reveals that the $r^2$ and *WTMM* methods identify as a continuum seizures that the *STA/LTA* and validated algorithms detect as clusters of short seizures. The lack of correspondence between a certain percentage of detections (11.8% for the $r^2$ method, 2.9% for the *WTMM* method and 8.5% for the *STA/LTA* method) and the validated algorithm may be partially attributed to brief discontinuities in seizure activity as shown in figure 5. This phenomenon ("go-stop-go") appears to be inherent to seizures (e.g., it is a general feature of intermittency associated with many dynamical systems). These discontinuities are also an "artifact" caused by the architecture of and parameters used in each algorithm. For example, the longer the foreground window and the higher the order statistical filter (e.g., median vs. quartile), in the validated algorithm, the higher the probability that "gaps" in seizure activity will occur.

Clustering of detections [6] is a strategy to manage dynamical or artifactual ictal "fragmentation".

The dependencies of seizure energy on seizure duration, for the set of seizures detected by each of the methods, are depicted in figure 7. A subset of seizures detected by all methods obeys a simple law of proportionality between energy and duration, that is, the longer the seizure, the largest its energy. However, this relationship is far from being invariably linear, indicating the presence of interesting scaling properties of seizure energy. Indeed, with the exception of the validated method, the others detect sets of seizures that are characterized by non-trivial scaling properties and much more variability in the standard deviation of the power of cortical activity. This can be surmised from the slopes being different from 1 (Figure 7) of the lower envelops of the scatter of points in panels (c) and (d) corresponding to the $r^2$ and *WTMM* methods, and to the nonlinear dark crescent seen in panel (b) corresponding to the *STA/LTA* method. The seizures detected by the validated algorithm have the smallest dispersion in the energy-duration relation.

The conditional probabilities of durations (figure 8a) and of the logarithm of energy of seizures (figure 8b) provide additional support that their properties are partly a function of the method used for their detection. The validated and *STA/LTA* algorithms yield similar durations but different from those of the *WTMM* and $r^2$ methods, which are analogous to each other (figure 8a). The distributions of the logarithm of seizure energies as identified by each of the methods (figure 8b) reveals additional discrepancies as evidenced by the much narrower and shorter "tail" distribution of the validated algorithm compared to the others.

To conclude, each of the investigated methods is "sensitive" to different seizure properties or features and may be regarded as providing complementary dynamical and clinical relevant knowledge with translational value. The *AIF* and *PIF* may be viewed as a first attempt towards a more nuanced definition (probabilistic) of seizures with operational value. That concordance levels between methods fluctuates as a function of seizure duration, commonly reaching its highest possible value (AIF=PIF=1) sometime (20-30s.) after onset, insinuate a decline in signal complexity or in its entropy, as feature homogeneity transitorily prevails over heterogeneity.

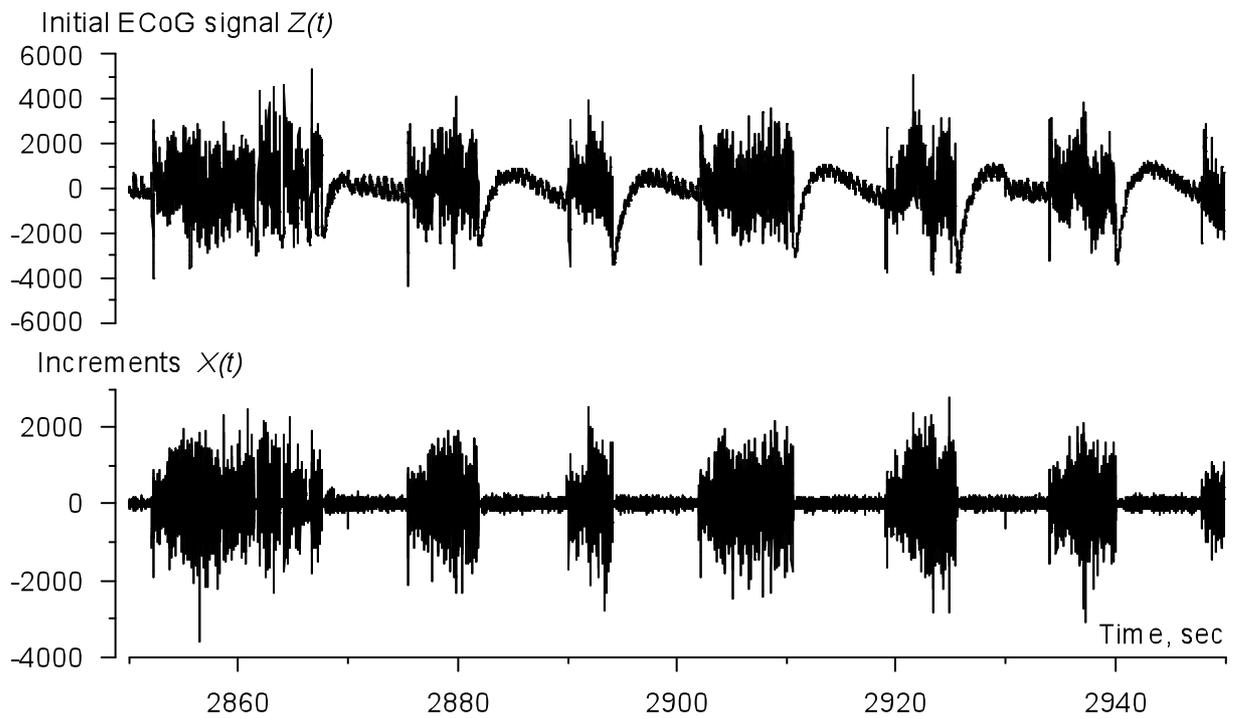

**Figure 1.** ECoG before (upper panel) and after differentiation (lower panel). The differentiated ECoG is less non-stationary (chiefly at low frequencies) than the undifferentiated one (*x*-axis: time in sec.; *y*-axis: amplitude in microvolts).

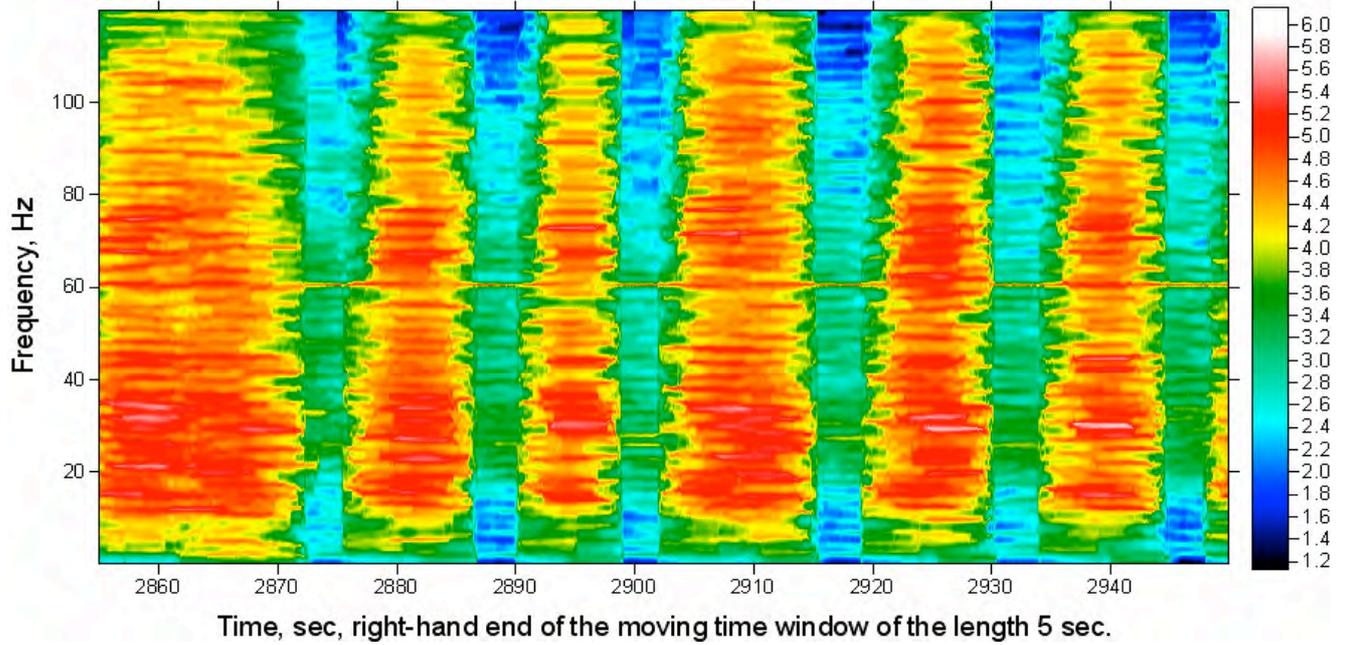

**Figure 2.** Temporal evolution of the decimal logarithm of the power spectrum of differentiated ECoG (as shown in Figure 1, bottom panel) estimated in 5s. moving windows. Six brief seizures appear as marked power spectrum increases (red and specks of white) in the 10-100 Hz. band (*x*-axis: time in sec.; *y*-axis: frequency (Hz); color scale to the right of main graph).

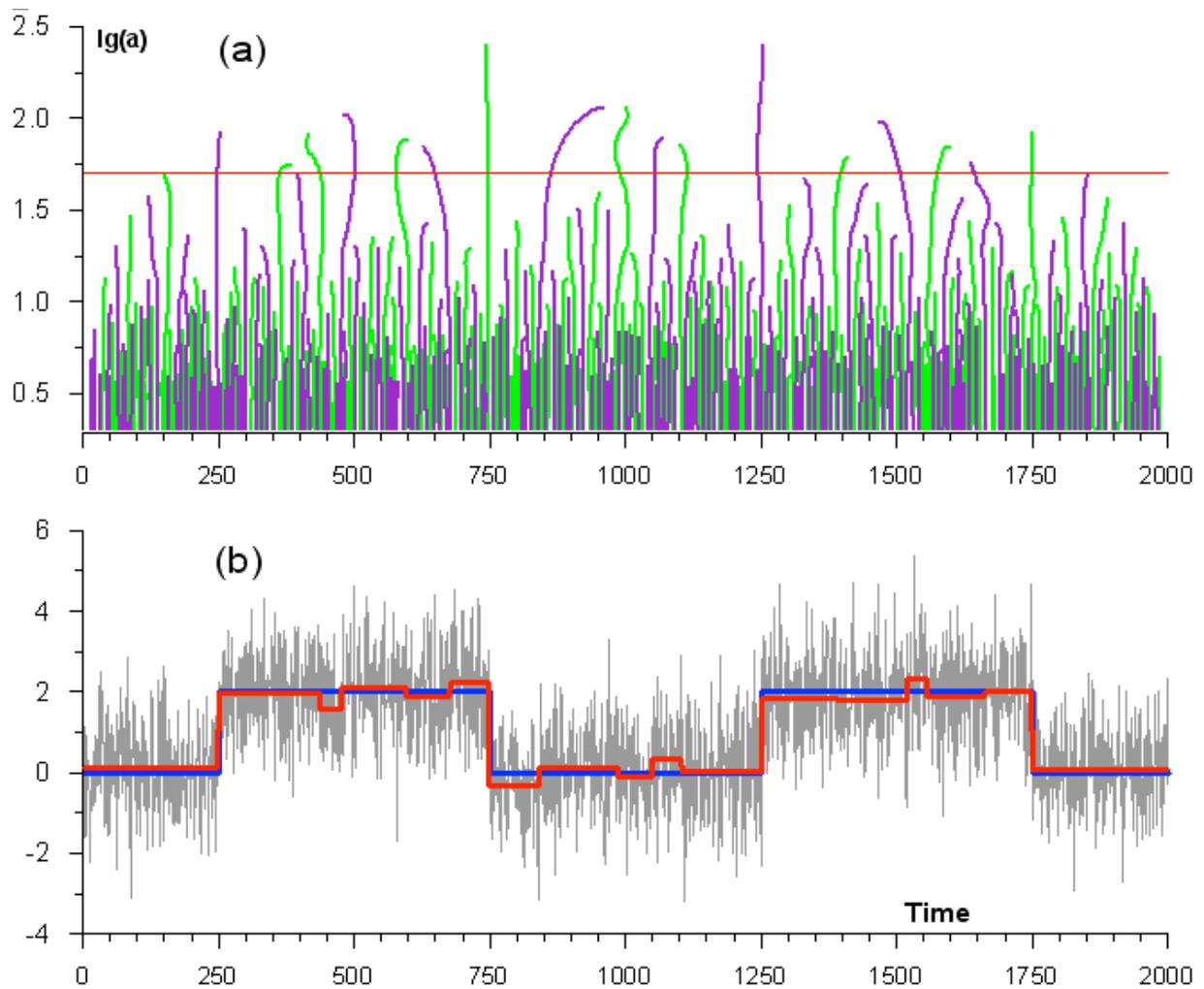

Figure 3. (For both panels, the x-axis represent time (in sec)). (a) y-axis: Wavelet Transform Maximum Modulus-chains constructed using the 1st derivative of the Gaussian wavelet for the signal $S_0(t) + \varepsilon(t)$ defined by expression (12) (grey lines in Fig. 3b). The green lines correspond to WTMM-chains for negative $c_1(t,a)$ values and the violet lines to WTMM-chains for positive $c_1(t,a)$ values. The thin red horizontal line indicates the time scale threshold $a_* = 50$; (b) y-axis: The grey lines represent the signal $S_0(t) + \varepsilon(t)$ (equation (12)), the bold blue line is the pure signal $S_0(t)$ without noise and the bold red line the reconstructed signal $S_Y(t \mid a_*)$ using the WTMM method for $a_* = 50$.

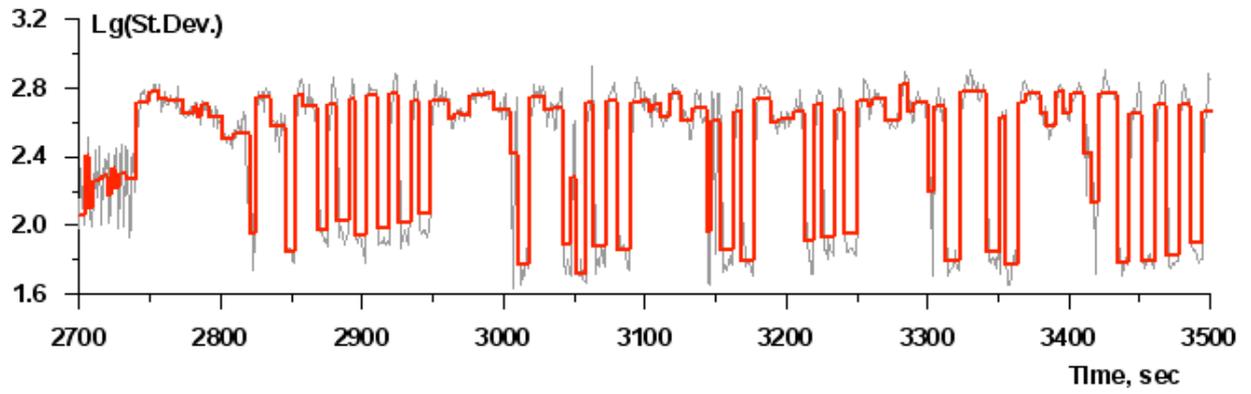

Fig.4. Grey line – values of the logarithmic standard deviation of the time series Rat48 increments, estimated within adjacent time intervals of length $L = 240$. Bold red line – its WTMM-stepwise approximation with time scale threshold $a_* = 3$.

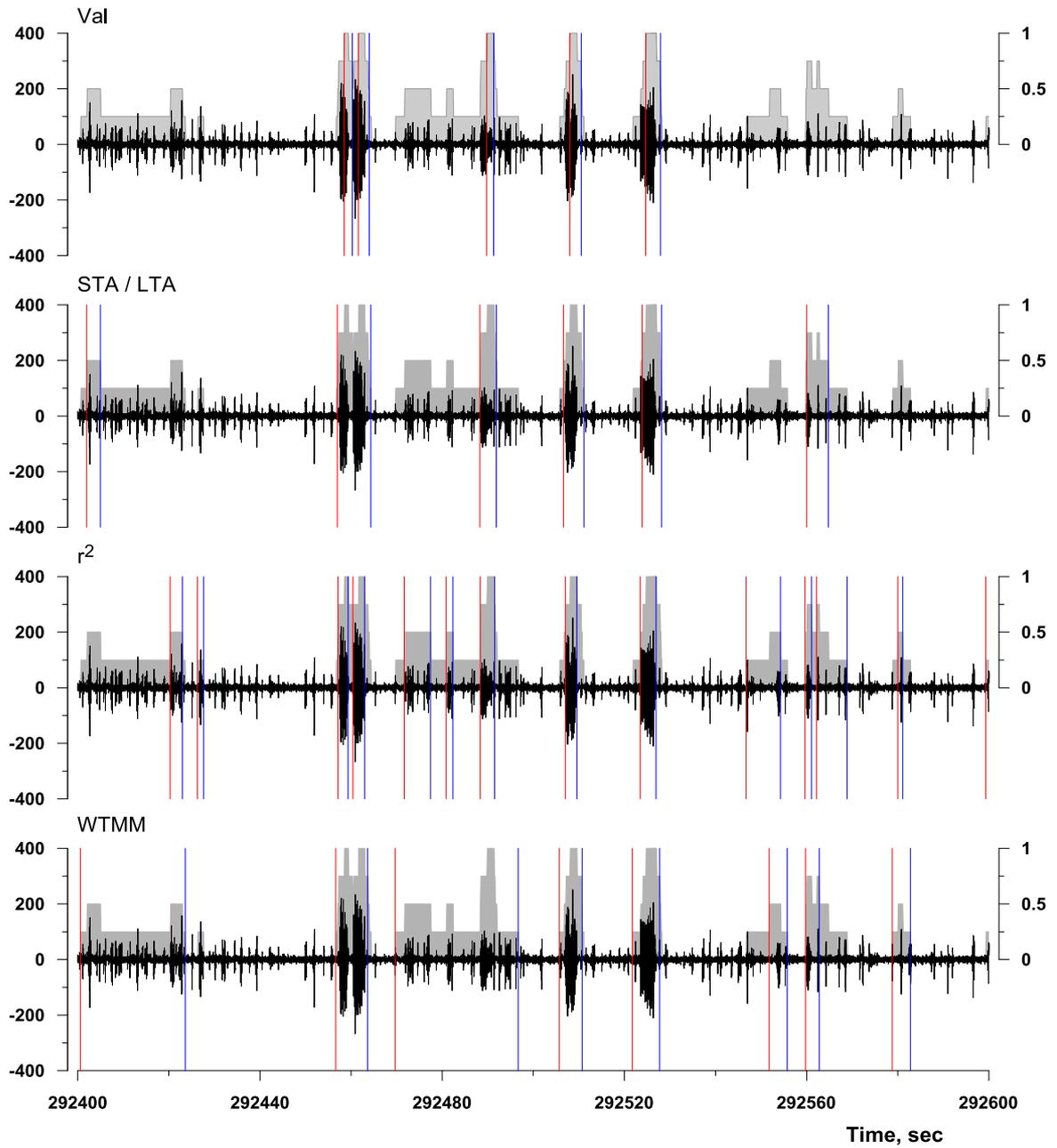

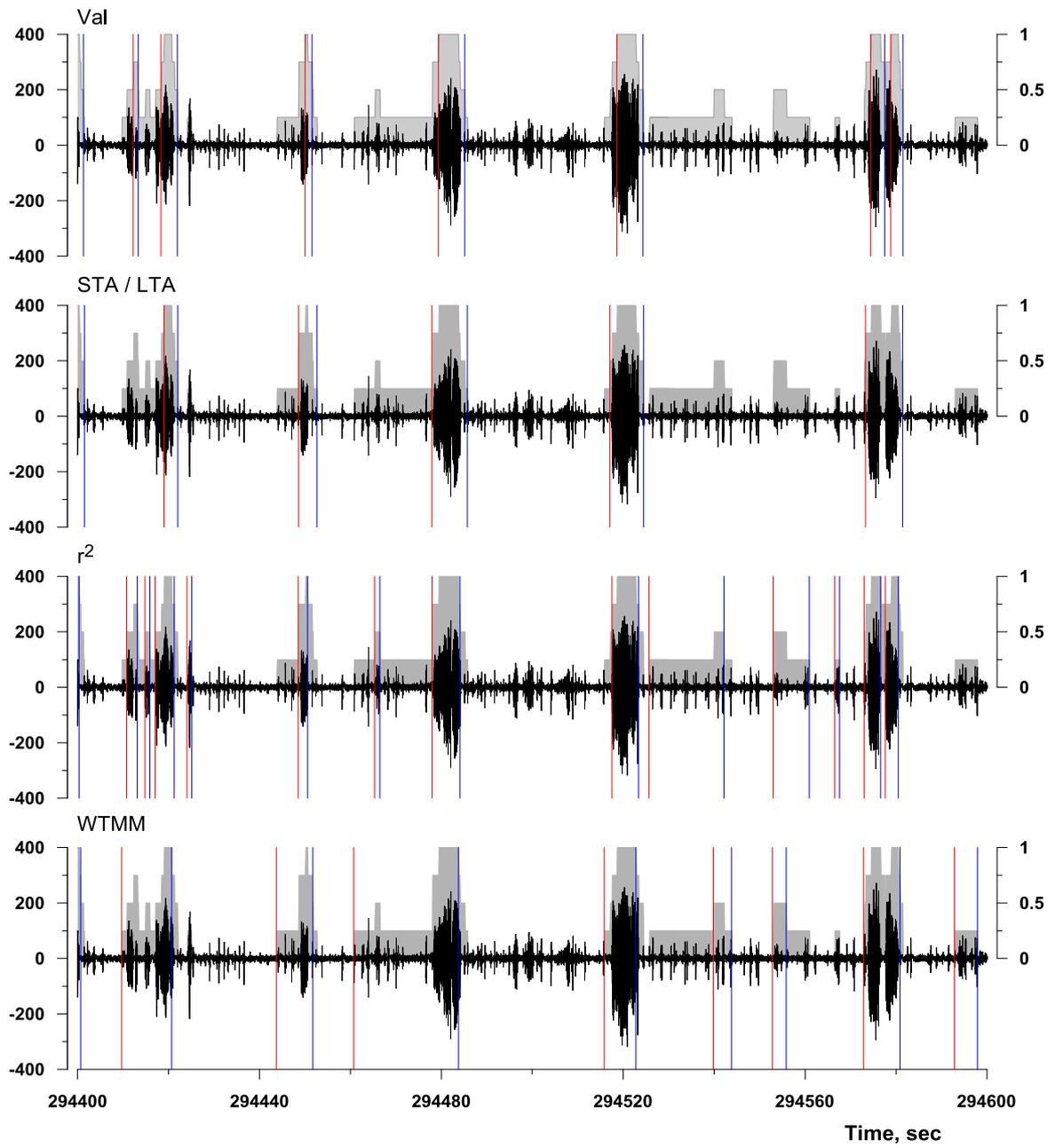

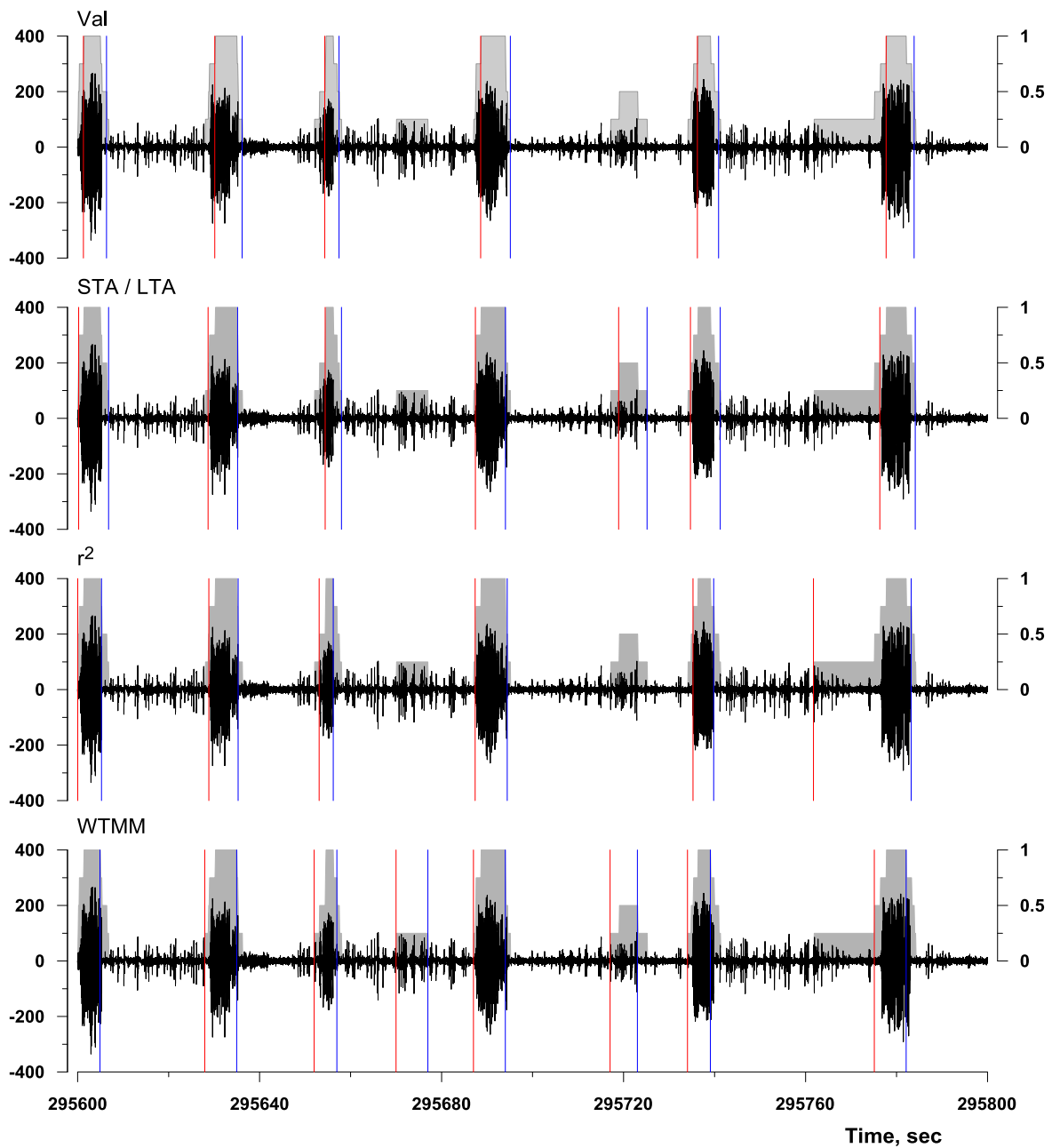

**Figure 5a,b,c.** Results of applying four different seizure detection methods (*Val*, *STA/LTA*, *r²*, *WTMM*) to the differentiated ECoG (in black; 200 s./panel) of a human with pharmaco-resistant epilepsy. The grey boxes represent the values (right *y*-axis) of the *Average Indicator Function* in the interval [0,1]. Seizures onset times are indicated by vertical red lines, and end times by vertical blue lines. Notice that the value of the *Average Indicator Function* is rarely *1*, at onset or termination, indicating **all** methods do **not** detect the ECoG activity as being **ictal** in nature. Left *y*-axis: ECoG amplitude (in µV); excursions above zero correspond to positive and below it, to negative polarity.

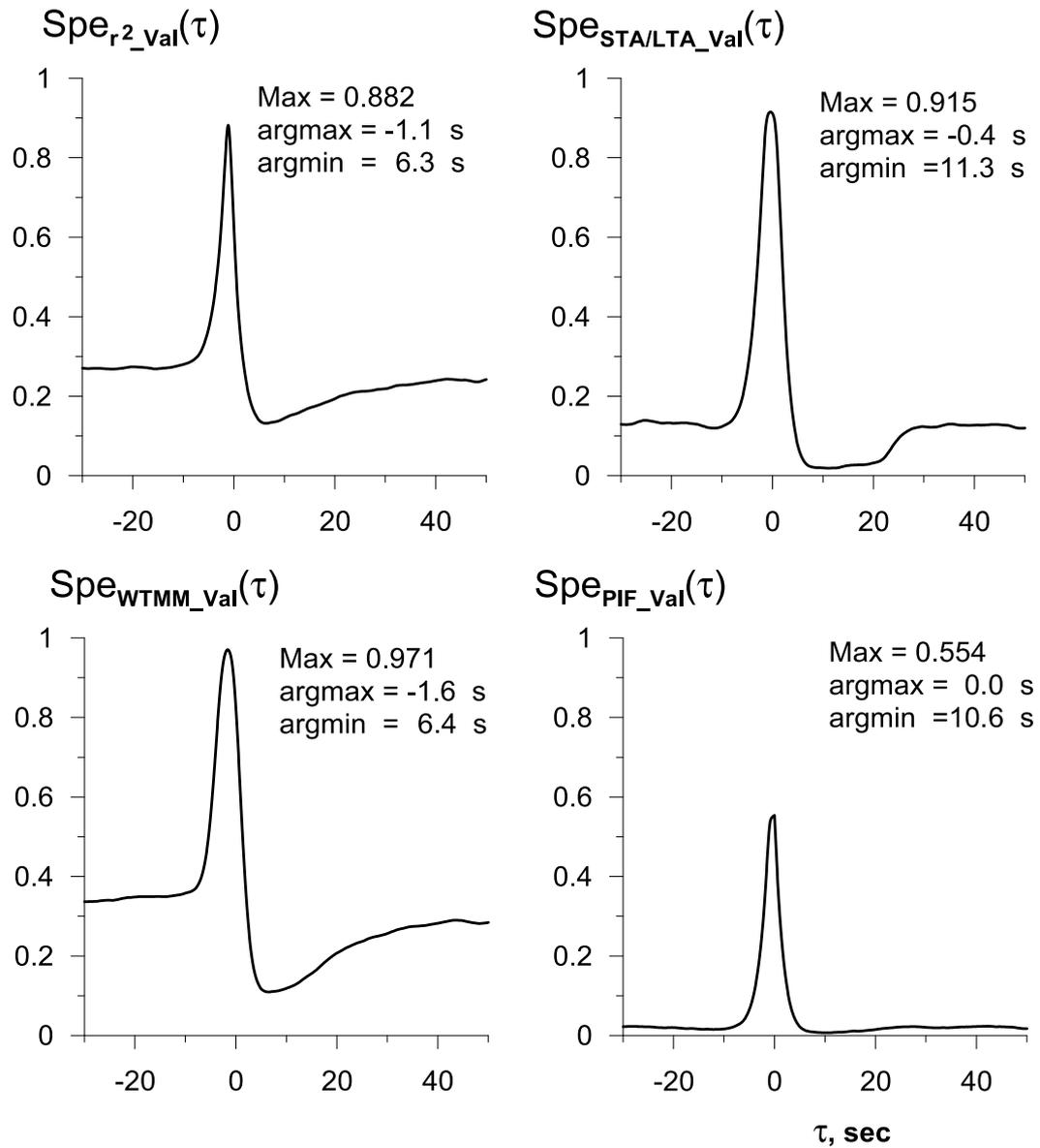

**Figure 6.** Graphics of specificity functions for each method as a function of time with respect to the Validated algorithms's time of seizure detection. *Upper left panel*: Auto-regressive model vs. Validated algorithm; *Upper right panel*: Short/Long Term Average Method vs. Validated algorithm; *Lower left panel*: Wavelet transform Maximum Modulus vs. Validated algorithm; *Lower right panel*: Product Index Function vs. Validated algorithm; only 55% of seizures detected by all methods are detected by the validated algorithm (*Val*). Tau (τ) zero (*x*-axis) corresponds to the time at which *Val* issues a detection. Negative τ values indicate "late" detections by the validated algorithm in relation to the other three and positive value the opposite. As shown above, $r^2$, *STA/LTA* and *WTMM* issue earlier detections than *Val.* Values of the lags τ corresponding to the maximum and minimum values of each function are presented for each graphic under the names argmax and argmin respectively.

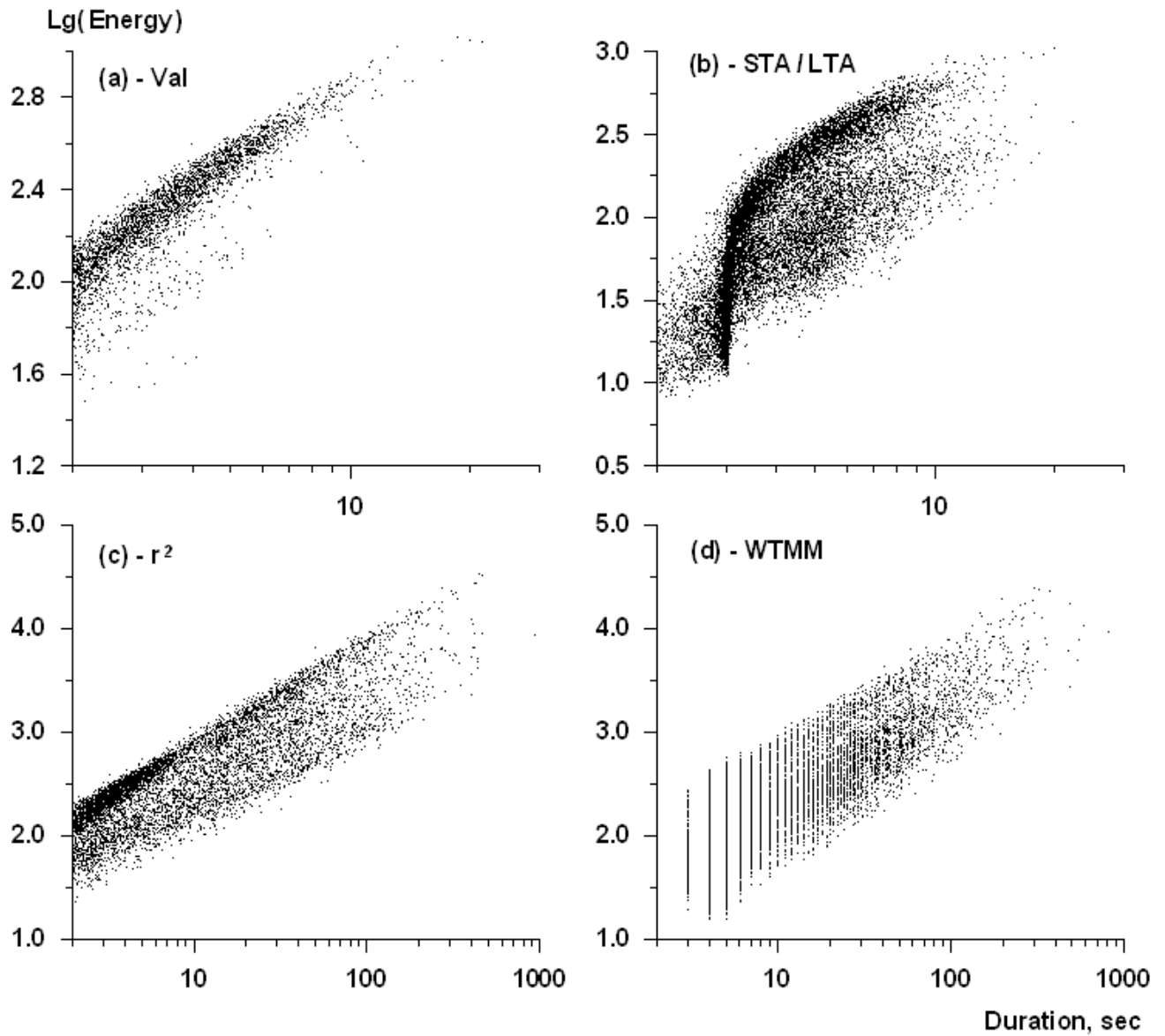

**Figure 7.** Plots of the decimal logarithm of the dependence of seizure energy on seizure duration (minimum duration: 2 sec.). Seizure is defined as the product of the standard deviation of the differentiated the ECoG and seizure duration (in sec.). *Upper left plot*: Validated algorithm detections; *Right upper plot*: Short/Long Term Average detections; *Left lower plot*: Auto-regressive model detections; *Right lower plot*: Wavelet-Transform Maximum Modulus detections. As expected, the differences in seizure onset and termination times are reflected in the energy-duration distributions; the dispersion of standard deviations varies widely among the different methods and non-linearities are present in certain distributions (e.g., panel b)).

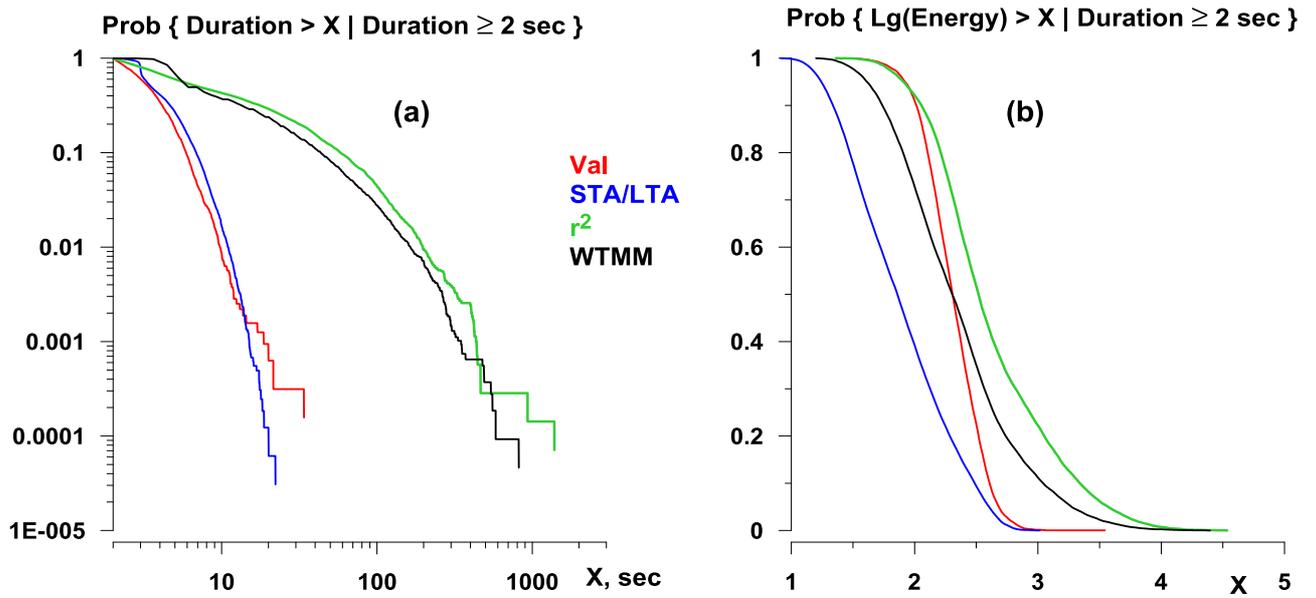

**Figure 8.** Empirical "tail" of the conditional probability distribution functions for: (a) Seizure durations (minimum duration: 2 sec); (b) the logarithm of seizure energy as estimated with the four different methods (Validated: Red; Short/Long Term Average: Blue; Auto-regressive model: Green; Wavelet Transform Maximum Modulus: Black).

| | Validated algorithm | $r^2$ | STA/LTA | WTMM |
|---|---|---|---|---|
| Total number of seizures with duration $\geq 2$ s. | 3184 | 7029 | 16275 | 10795 |
| Mean duration, s. | 3.8 | 23 | 4.3 | 18.6 |
| Median duration, s. | 3.4 | 7 | 3.5 | 6 |
| % time spent in seizure | 2 | 27 | 12 | 34 |

Table 1. Summary statistics obtained by applying four different detection methods (Validated Algorithm; $r^2$; STA/LTA; WTMM) to a prolonged human seizure time-series. The minimum duration of seizures was set at 2 s because such duration is the minimum possible for the WTMM method with the parameter *L=240*.

| Method | $Spe_{Method\_Val}(0)$ | $\max_{\tau} Spe_{Method\_Val}(\tau)$ | $\arg\max_{\tau} Spe_{Method\_Val}(\tau)$ |
|---|---|---|---|
| $r^2$ | 0.628 | 0.882 | -1.1 s |
| *WTMM* | 0.823 | 0.971 | -1.6 s |
| *STA/ LTA* | 0.911 | 0.915 | -0.4 s |

Table 2. Values of specificity of the three novel methods calculated with respect to the validated method and time lag (as defined in the text) at which the specificities attain their largest values.